\documentclass{article}

\usepackage{arxiv}
\usepackage{url}
\usepackage{ascmac}
\usepackage[T1]{fontenc}    
\usepackage{hyperref}       
\usepackage{url}            
\usepackage{booktabs}       
\usepackage{amsfonts}       
\usepackage{nicefrac}       
\usepackage{microtype}      
\usepackage{lipsum}		
\usepackage[dvipdfmx]{graphicx}
\usepackage{color}

\usepackage{doi}
\usepackage{pdfpages}
\usepackage{multicol}
\usepackage{regexpatch,xparse}
 \usepackage[
    backend=biber,
    style=numeric,
  ]{biblatex}

\makeatletter

\let\do@footnotetext\@footnotetext
\regexpatchcmd{\do@footnotetext}
  {\c{insert}\c{footins}\cB.(.*)\cE.}
  {\1\c{egroup}}
  {}{}
\def\@footnotetext{\insert\footins\bgroup\@makeother\#\do@footnotetext}
\newenvironment{figurehere}
  {\def\@captype{figure}}
  {}
\newenvironment{tablehere}
  {\def\@captype{table}}
  {}
\newcommand{\ttvar}{\begingroup\@makeother\#\@ttvar}
\newcommand{\@ttvar}[1]{\ttfamily\detokenize{#1}\endgroup}
\makeatother

\title{Extracting user needs with Chat-GPT for dialogue recommendation}

\author{ 
	Yugen Sato\\
	Meiji University\\
	Kanagawa, Japan\\
	\texttt{yugen\_sato@cs.meiji.ac.jp} \\
	\And
	Taisei Nakajima\\
	Meiji University\\
	Kanagawa, Japan\\
        \And
	Tatsuki Kawamoto\\
	Meiji University\\
	Kanagawa, Japan\\
        \And
	Tomohiro Takagi\\
	Meiji University\\
	Kanagawa, Japan\\
 }

\renewcommand{\shorttitle}{\textit{arXiv} Template}

\hypersetup{
pdftitle={Extracting user needs with Chat-GPT for dialogue recommendation},
pdfsubject={q-bio.NC, q-bio.QM},
pdfauthor={David S.~Hippocampus, Elias D.~Striatum},
pdfkeywords={First keyword, Second keyword, More},
}

\begin{document}
\maketitle

\begin{abstract}
Large-scale language models (LLMs), such as ChatGPT, are becoming increasingly sophisticated and exhibit human-like capabilities, playing an essential role in assisting humans in a variety of everyday tasks. An important application of AI is interactive recommendation systems that respond to human inquiries and make recommendations tailored to the user. In most conventional interactive recommendation systems, the language model is used only as a dialogue model, and there is a separate recommendation system. This is due to the fact that the language model used as a dialogue system does not have the capability to serve as a recommendation system.
Therefore, we will realize the construction of a dialogue system with recommendation capability by using OpenAI's Chat-GPT, which has a very high inference capability as a dialogue system and the ability to generate high-quality sentences, and verify the effectiveness of the system.
\end{abstract}


\section{Introduction}
Large-scale language models (LLMs) such as ChatGPT \cite{chatgpt} have shown remarkable performance in a variety of natural language processing (NLP) tasks. By leveraging large-scale pre-training on large text corpora and reinforcement learning from human feedback (RLHF), LLMs not only have extensive knowledge, but also perform well in language understanding, generation, interaction, and inference.
GPT-4 \cite{OpenAI2023GPT4TR} even exceeds human performance. Prompt engineering techniques (chain-of-thought \cite{wei2023chainofthought} and in-context learning) allow LLMs to unlock their unlimited potential to perform the complex tasks of everyday life, making LLMs the subject of significant attention from both academia and industry. ChatGPT is a successful example of such an application, where the AI model is equipped with the ability to analyze context and respond to user queries based on knowledge gained from vast amounts of training data. It has the ability to respond to user queries. \\
In traditional interactive recommendation systems, the language model is used only as a dialogue model in many cases, and the recommendation system exists separately from it. The system provided recommendations to the user only when the dialogue model and the recommendation system interacted with each other.
The reason for this is that the dialogue system does not have the capability to act as a recommendation system, and OpenAI's Chat-GPT, with its very high inference capability and ability to generate high quality sentences, can make recommendations in a dialogue.
Therefore, we will experiment with OpenAI's Chat-GPT as a dialogue system with recommendation capability.
In a typical interactive recommendation system, the user asks the system a question such as "I am looking for an item like . and the system responds with some kind of recommendation. However, since our goal is to extract the user's needs, we assume a situation where the system asks the user questions to analyze the user's preferences and thoughts.
We will test the effectiveness and ability of Chat-GPT to dynamically manage the flow of dialogue in such a situation.

\section{Related works}
\subsection{Task-Oriented dialogue without a large language model}
Yichi Zhang et al. \cite{Zhang_Ou_Yu_2020}introduce a novel data augmentation approach and model architecture for generating multiple appropriate responses in task-oriented dialogue systems. The main contribution is to exploit the one-to-many mapping between dialogue states and valid system actions.
Focusing on task-oriented dialogue systems that can generate multiple appropriate responses under the same context, we proposed a Multi-Action Data Augmentation (MADA) framework MADA uses training data to generate a one-to-many mapping from dialogue states to valid system actions by MADA is trained to discover all possible mappings from dialogue states to valid system actions, thereby enabling it to generate diverse and appropriate dialogue responses.
\subsection{Task-Oriented dialogue}
Lang Cao \cite{cao2023diaggpt} introduces a new method called Dialogue in Diagnosis GPT (DiagGPT) that extends large-scale language models (LLMs) to task-oriented dialogue (TOD) scenarios. The main focus is to improve dialogue in complex diagnostic scenarios such as legal and medical consultations where simple question-answer dialogue is not sufficient. The approach aims to guide the user toward specific task completion, which is a hallmark of TOD, by having the AI chat agent proactively ask questions.\\
Task Guidance: Guides the user to a specific goal by sequencing predefined topics and assists in accomplishing the task through the progression of the dialogue.\\
Proactive Questioning: Gather necessary information from the user by proactively asking questions based on a predefined checklist.\\
Topic Management: Automatically manage topics in dialogue, track topic progression, and effectively participate in discussions around the current topic.\\
Highly scalable: DiagGPT is designed to be flexible enough to incorporate additional features to handle tasks in complex scenarios or to accommodate more needs in the conversation system.\\
DiagGPT is presented as a multi-agent AI system with automatic topic management capabilities to enhance its usefulness in task-oriented dialogue scenarios. Through this design, we aim to better simulate real medical and legal professionals and provide a more intelligent and professional chatbot experience.
In the meantime, we will test whether this topic management and task guiding functionality can be provided by a single module, the controller(\ref{3.1}).

\section{Method}
\label{3}
Our framework consists of three modules, "Controller" "Assistant" and "User Simulator" and is outlined in Figure \ref{fig:1}. This chapter describes the structure of the framework, the role of each module, and the relationship between them.
\begin{multicols}{2}
\subsection{Controller and Assistant}
\label{3.1}
This subsection describes two of the three elements: the controller and the assistant.
First, the assistant is responsible for interacting directly with the user, generating questions for the user that elicit user characteristics and needs. The controller does not interact with the user, but is responsible for monitoring the assistant's speech and controlling the interaction between the user and the assistant from a higher level.
The flow of the system is as follows As shown in Figure \ref{fig:1}, the query generated by the assistant is passed to the controller, who determines whether the content of the query is appropriate in terms of the dialogue scenario and communicates it to the assistant.
\columnbreak
\columnbreak
\begin{figurehere}
    \includegraphics[width=\columnwidth]{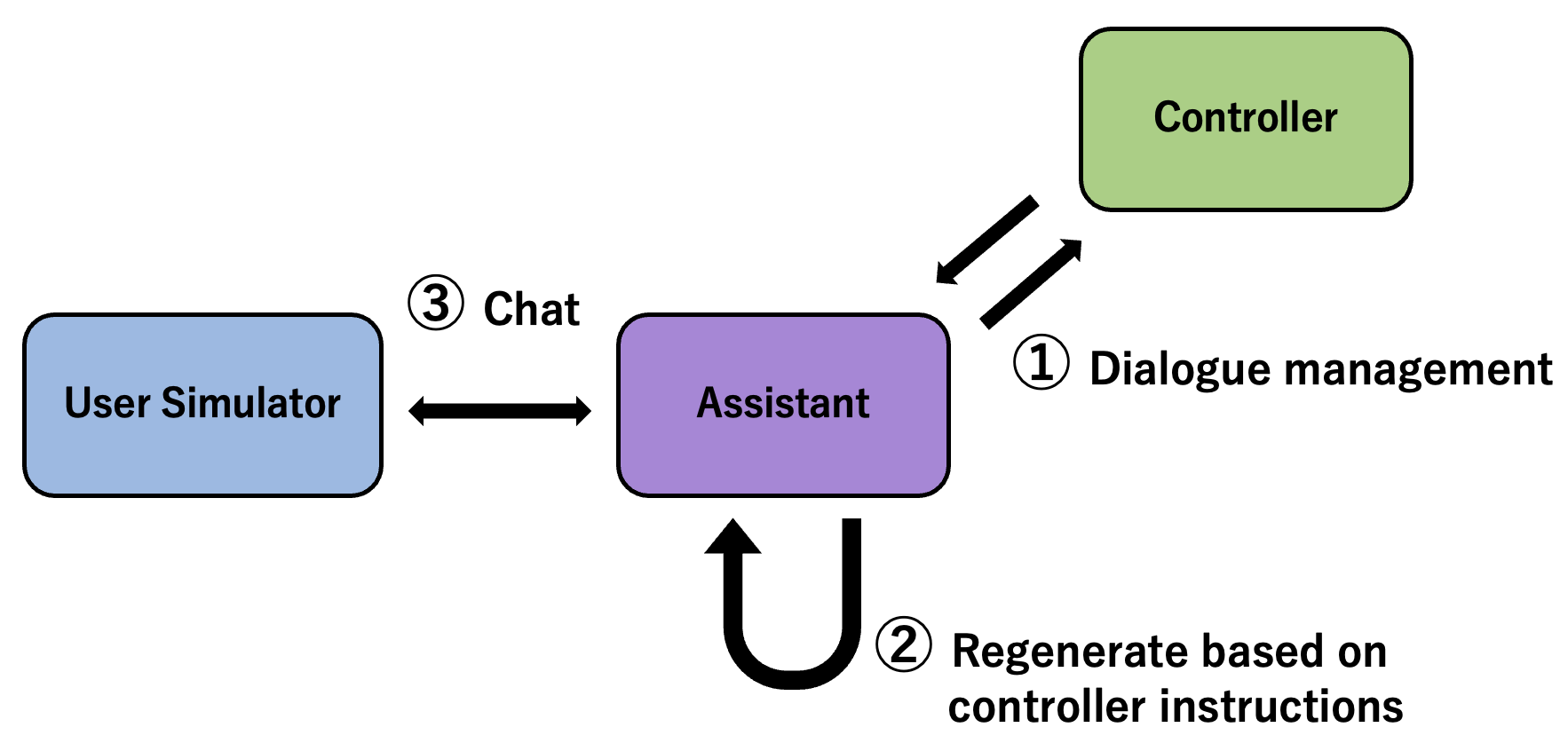}
     \caption{Architecture of the system}
     \label{fig:1}
\end{figurehere}
\end{multicols}
In Chat-GPT, the dialogue schema is predetermined, and the flow of dialogue can be configured according to the schema by inputting it into the prompts. However, this alone will fix the content of the dialogue and lead to a decrease in user satisfaction in terms of personalization of the dialogue. Therefore, instead of a hard constraint such as specifying the schema in the prompt, we can change it to a soft constraint such as instructions on question generation to the dialog module by the dialog management module, thereby allowing more freedom in the content of the dialog and realizing dynamic dialog.
The following are prompts from the controller and assistant on the subject of restaurants recommendation.
\begin{itembox}[l]{Controller Prompt (Examples of Restaurant Recommendations)}
You are the controller who controls the conversation between the Assistant and the user for restaurants recommendation. The assistant will, at your direction, come up with the next question to ask the user.\\\\
\ttvar{###Input Information and Dialogue Termination Conditions}\\
The input you are given is the question generated by the assistant and the user's answer to it, plus the statement "Terminate an assistant's recommendation?" statement. If the amount of information elicited by the dialog control is deemed sufficient, it is necessary to generate instructions for the assistant to communicate the user's needs after the "Yes".
Otherwise, just output "No".\\
\\
The questions that should be created depend on the case, but it is good to consider questions with the background of eliciting the potential needs of the user. Initially, it is best to focus on the user's daily life and get basic information before analyzing their needs.\\
Then, for example, topics could include "purpose/use of the restaurant," "with whom to use the restaurant," "conditions to avoid," "budget," etc.  These are only examples, and may not be correct, so please use them as a reference to structure the flow of dialogue for recommending properties yourself, and create instructions for the assistant.\\
If there are other questions that should be asked during the conversation with the user, please actively generate new questions.\\
\\
Now generate an instruction that asks the user to think of an initial question.\\
\\
\ttvar{###NOTES}\\
The instructions you generate are not specific instructions, but rather they provide the axis around which the assistant will think about the question.\\
Your generated content will be conveyed verbatim to the assistant, so please omit response phrases such as "I understand.
Please generate the form following "\ttvar{###Controller}".
\end{itembox}
\\
\begin{itembox}[l]{Assistant Prompt (Examples of Restaurant Recommendations)}
You are an assistant who proposes restaurants based on conversations with users. The conversation with the user is controlled by the controller. You are required to follow the controller's instructions to converse with the user. The controller's instructions are input as needed. If there is no instruction, please think of the best question to ask the user by yourself considering the conversation up to that point.\\\\
\ttvar{###NOTES}\\
- Output should be in the format following "\ttvar{###Assistant}" and generate direct questions only.\\
- Only the role of the assistant should be performed.\\
- Do not make any evaluation predictions until instructed to do so by the controller.\\
- Do not generate controller statements.\\
Now follow the controller's instructions to extract user preferences as an assistant.\\\\
\ttvar{###Controller}\\
\ttvar{{controller_instruction}}\\\\
\end{itembox}
\newpage
\subsection{User simulator}
\label{3.2}
In making our system work, how we implement user utterances is very important. In the case of general interactive recommendation, user speech is often implemented using a publicly available dialogue recommendation dataset. Since the data contained in a dialogue recommendation dataset is a pair of a question by the user and a response by the system, and there is no pair data of a question by the system and a response by the user, it is difficult to experiment with a public dataset. Therefore, we attempt to solve this problem by simulating users using Chat-GPT.
We have created a module that incorporates the user's gender, age, and occupation into the prompts and outputs the user's responses, which act as input to the assistant's utterances. This allows us to validate a dynamic dialogue system between the assistant and the controller.
Below are the prompts for the user simulator on the subject of restaurants recommendation as well as the controller and assistant.
\begin{itembox}[l]{User Simulator Prompt (Examples of Restaurant Recommendations)}
As I enter the Assistant's remarks, which is a Chat Bot, you generate a conversation in which you, as the user, respond to the Assistant.\\
The assistant will conduct a needs analysis for restaurants recommendations based on your statements.\\\\
\ttvar{###User Information you will play}\\
- Gender: Male\\
- Age: 24\\
- Occupation: Engineer\\
- Favorite Cuisine: Italian\\
- Occasion: Company get-togethers\\\\
Based on the user information you play, please have a conversation with me, your assistant. Please begin your output with "\ttvar{###User}.\\\\
\ttvar{###NOTES}\\
- Please make statements that are sometimes contradictory. This will help the assistant determine if the user's inconsistencies can be adequately addressed.\\
- Output should be user statements only.\\
- Please try to use randomness in your speech, such as answering not only the question asked but also other things as well.\\
\\
\ttvar{###Assistant}\\
\ttvar{{assistant_context}}
\end{itembox}

\newpage

\begin{multicols}{2}
    \section{Experiment}
    \label{4}
    In this section, we experiment with the system described in \ref{3} on the subject of recommending restaurants.
    The assistant's main objective is to repeat various questions to a user looking for a restaurant to extract the user's needs, and the controller's objective is to support that process.
    In generating the user simulator, we do not use a specific dialogue data set, but only prior knowledge of the user simulator to conduct the dialogue with the assistant.
    The model used in the experiments was GPT-4 for the assistant, controller, and user simulator, all of which were generated using OpenAI's web-based module, \cite{chatgpt-web}. The prompts for each module are the same as those introduced in \ref{3.1} and \ref{3.2}, with the generated content of each module inserted in the area enclosed in \ttvar{{}}.
    An example of the interaction between the user simulator and the assistant is shown in Figure 1.
    \subsection{Qualitative evaluation}
    Figure \ref{fig:2} shows an example of dialogue in restaurants recommendation. The user is looking for an restaurant that meets his/her requirements, and the system needs to help him/her to do so.
    The development of the dialogue is dynamically controlled by the controller. The dialogue includes the "purpose and use of the restaurant," "who will use it with," "budget," etc., which are also clearly indicated in the prompts.In addition, "priorities" and "schedule confirmations" are also generated under the controller's control.
    Due to space limitations, we are unable to present further examples of dialogue in this paper. Our system simulates many real consulting scenarios.
    
    \subsection{Evaluation by user simulator}
    In this section, we use evaluation by the user simulator itself as a method of dialogue evaluation.
    G-EVAL \cite{liu2023geval} proposed a method for evaluating natural language generation tasks using GPT-4, and it was shown that evaluation using GPT-4 is the closest to human evaluation compared to conventional methods. G-EVAL also introduces evaluation of dialogue generation tasks, which we apply to evaluate whether the entire dialogue is well controlled by the system.
    Specifically, we input the dialogue history as shown in Figure \ref{fig:2}and ask the user to score each of the following four items on a scale of 1 to 5.
    \begin{itemize}
        \setlength{\itemindent}{-10mm}
        \item satisfaction
        \begin{itemize}
            \setlength{\itemindent}{-10mm}
            \item Whether the user was satisfied with the dialogue
        \end{itemize}
        \item flexibility
        \begin{itemize}
            \setlength{\itemindent}{-10mm}
            \item Whether you were able to compose a tactful flow of dialogue based on the user's statements
        \end{itemize}
        \item accuracy
        \begin{itemize}
            \setlength{\itemindent}{-10mm}
            \item Whether we were able to accurately identify user needs and organize the information
        \end{itemize}
    \end{itemize}
    \columnbreak
    
    \begin{figurehere}
    \includegraphics[width=0.5\textwidth,height=21cm]{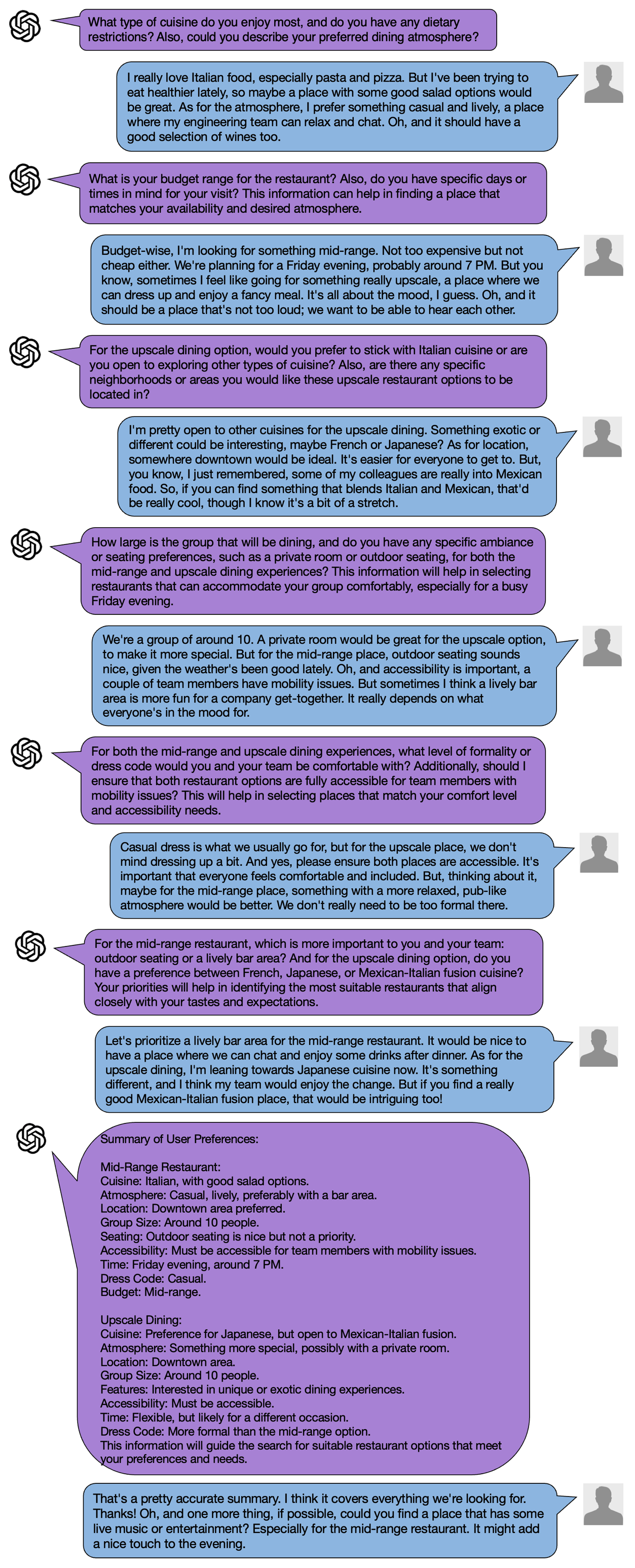}
    \caption{Example of user-assistant interaction in recommending restaurants}
    \label{fig:2}
    \end{figurehere}
\end{multicols}

\newpage
\begin{tablehere}
    \centering
    \begin{tabular}{ccccc}
        \toprule
         & satisfaction & flexibility & accuracy & contradiction\\
        \midrule
       Ours & 4.7 & 4.9  & 4.8 & 4.2\\
       \midrule
       GPT-4 & 4.2 & 3.9 & 4.1 & 3.0\\
       \bottomrule
    \end{tabular}
    \caption{Evaluation by user simulator}
    \label{tab:1}
    \end{tablehere}
    
\begin{multicols}{2}
\begin{itemize}
    \setlength{\itemindent}{-10mm}
    \item contradiction
    \begin{itemize}
        \setlength{\itemindent}{-10mm}
        \item Whether you were able to successfully approach the user's statement by pointing out the inconsistencies hidden in the user's statement
    \end{itemize}
\end{itemize}
Our system and the GPT-4 each interact with a user simulator and are evaluated on the above four items.
Table 1 shows the comparative results of the average scores evaluated using the five dialogue data.
    \subsection{Interactive control by controller}
    An example of the controller's control of the assistant's behavior is shown in Figure \ref{fig:3}. As can be seen from this example, the controller can take the questions generated by the assistant to the user, determine whether they are correct for the final purpose and not unnatural for the flow of the dialogue, and apply the necessary modifications. The modifications are not specific, such as giving an answer to the assistant, but are made at a level of abstraction that indicates only the policy, so that the content of the dialogue can maintain a certain degree of freedom, rather than a fixed development. This makes it possible to construct a more personalized dialogue that fits the user's characteristics, and is thought to have improved satisfaction.
    \columnbreak
    
    \begin{figurehere}
     \includegraphics[width=\columnwidth]{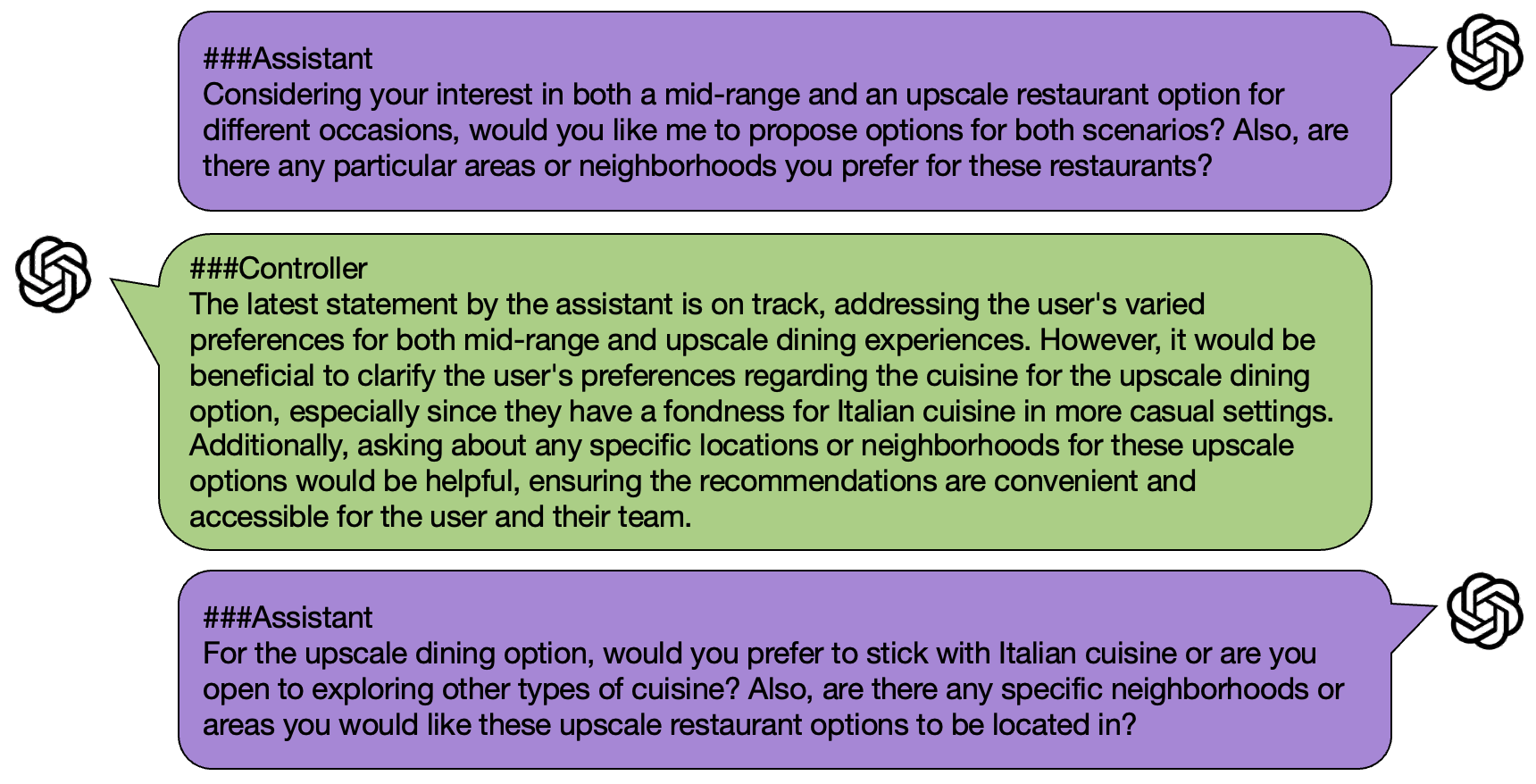}
     \caption{Example of controller interactive control and assistant generation modification}
     \label{fig:3}
    \end{figurehere}
\end{multicols}

\section{Conclusion}
We verified Chat-GPT's capabilities and effectiveness in managing dynamic dialogues for the purpose of extracting user needs by simulating dialogues between a controller and an assistant using Chat-GPT.
As a result, we confirmed that our system is capable of dynamically controlling the dialogue while inferring user preferences and requests, and forming dialogue scenarios that are highly satisfactory for the user.
We expect that Chat-GPT's dialogue management capabilities will be used in the future for applications that are truly valuable to users.
\printbibliography

@article{OpenAI2023GPT4TR,
  author = {OpenAI},
  description = {This paper reports the development of GPT-4, a large-scale, multimodal model which can accept image and text inputs and produce text outputs. GPT-4 exhibits human-level performance on various professional and academic benchmarks, including passing a simulated bar exam with a score around the top 10% of test takers.},
  journal = {ArXiv},
  title = {GPT-4 Technical Report},
  volume = {abs/2303.08774},
  year = 2023
}

@Misc{chatgpt-web,
  title        = "{Open AI,Chat-GPT}",
  howpublished = "\url{https://chat.openai.com/}"
}

@misc{cao2023diaggpt,
      title={DiagGPT: An LLM-based Chatbot with Automatic Topic Management for Task-Oriented Dialogue}, 
      author={Lang Cao},
      year={2023},
      eprint={2308.08043},
      archivePrefix={arXiv},
      primaryClass={cs.CL}
}

@misc{wei2023chainofthought,
      title={Chain-of-Thought Prompting Elicits Reasoning in Large Language Models}, 
      author={Jason Wei and Xuezhi Wang and Dale Schuurmans and Maarten Bosma and Brian Ichter and Fei Xia and Ed Chi and Quoc Le and Denny Zhou},
      year={2023},
      eprint={2201.11903},
      archivePrefix={arXiv},
      primaryClass={cs.CL}
}

@misc{liu2023geval,
      title={G-Eval: NLG Evaluation using GPT-4 with Better Human Alignment}, 
      author={Yang Liu and Dan Iter and Yichong Xu and Shuohang Wang and Ruochen Xu and Chenguang Zhu},
      year={2023},
      eprint={2303.16634},
      archivePrefix={arXiv},
      primaryClass={cs.CL}
}

@Misc{chatgpt,
  title        = "{Open AI,Chat-GPT}",
  howpublished = "\url{https://openai.com/blog/chatgpt}"
}

@article{Zhang_Ou_Yu_2020, title={Task-Oriented Dialog Systems That Consider Multiple Appropriate Responses under the Same Context}, volume={34}, url={https://ojs.aaai.org/index.php/AAAI/article/view/6507}, DOI={10.1609/aaai.v34i05.6507}, abstractNote={&lt;p&gt;Conversations have an intrinsic one-to-many property, which means that multiple responses can be appropriate for the same dialog context. In task-oriented dialogs, this property leads to different valid dialog policies towards task completion. However, none of the existing task-oriented dialog generation approaches takes this property into account. We propose a Multi-Action Data Augmentation (MADA) framework to utilize the one-to-many property to generate diverse appropriate dialog responses. Specifically, we first use dialog states to summarize the dialog history, and then discover all possible mappings from every dialog state to its different valid system actions. During dialog system training, we enable the current dialog state to map to all valid system actions discovered in the previous process to create additional state-action pairs. By incorporating these additional pairs, the dialog policy learns a balanced action distribution, which further guides the dialog model to generate diverse responses. Experimental results show that the proposed framework consistently improves dialog policy diversity, and results in improved response diversity and appropriateness. Our model obtains state-of-the-art results on MultiWOZ.&lt;/p&gt;}, number={05}, journal={Proceedings of the AAAI Conference on Artificial Intelligence}, author={Zhang, Yichi and Ou, Zhijian and Yu, Zhou}, year={2020}, month={Apr.}, pages={9604-9611} }
\end{document}